\documentclass[letterpaper,prl,floatfix,showpacs,twocolumn]{revtex4}
\usepackage{amssymb,amsmath,graphicx,xspace,overpic,ifthen}
\pdfoutput=1
\begin{document}

\def\Id{1\!\mathrm{l}}
\newcommand{\Tr}{\mathrm{Tr}}
\newcommand{\ket}[1]{\ensuremath{\left|\scriptstyle#1\right\rangle}}
\newcommand{\bra}[1]{\ensuremath{\left\langle\scriptstyle#1\right|}}
\newcommand{\braket}[2]{\ensuremath{\left\langle\scriptstyle#1|#2\right\rangle}}
\newcommand{\ketbra}[2]{\ket{#1}\!\!\bra{#2}}
\newcommand{\expect}[1]{\ensuremath{\left\langle#1\right\rangle}}
\newcommand{\braopket}[3]{\ensuremath{\bra{#1}#2\ket{#3}}}
\newcommand{\proj}[1]{\ketbra{#1}{#1}}

\newcommand{\diff}{\ensuremath{\mathrm{d}}}
\newcommand{\z}{\ensuremath{\vec{z}}}
\newcommand{\R}{\ensuremath{\mathbf{R}}}
\newcommand{\M}{\ensuremath{\mathbf{M}}}

\newcommand{\Sys}{\ensuremath{\mathcal{S}}}
\newcommand{\Env}{\ensuremath{\mathcal{E}}}
\newcommand{\Frag}{\ensuremath{\mathcal{F}}}

\newcommand{\rhoS}{\ensuremath{\rho_{\Sys}}}
\newcommand{\rhoE}{\ensuremath{\rho_{\Env}}}
\newcommand{\rhoSE}{\ensuremath{\rho_{\Sys\Env}}}
\newcommand{\rhoF}{\ensuremath{\rho_{\Frag}}}
\newcommand{\rhoSF}{\ensuremath{\rho_{\Sys\Frag}}}
\newcommand{\Nenv}{\ensuremath{N_{\mathrm{env}}}}
\newcommand{\idx}[1]{^{(#1)}}

\newcommand{\I}{\ensuremath{\mathcal{I}}}
\newcommand{\Ise}{\ensuremath{\I_{\Sys:\Env}}}
\newcommand{\Isf}{\ensuremath{\I_{\Sys:\Frag}}}
\newcommand{\Isen}[1]{\ensuremath{\I_{\Sys:\Env_{#1}}}}
\newcommand{\Ibar}{\ensuremath{\overline{\I}}}

\newcommand{\Hh}{\ensuremath{H}}
\newcommand{\Hs}{\ensuremath{\Hh_{\Sys}}}
\newcommand{\Hmax}{\ensuremath{\Hh_{\mathrm max}}}
\newcommand{\He}[1][]{\ensuremath{\Hh_{\Env_{#1}}}}
\newcommand{\Hf}{\ensuremath{\Hh_{\Frag}}}
\newcommand{\Hse}[1][]{\ensuremath{\Hh_{\Sys\Env_{#1}}}}
\newcommand{\Hsf}{\ensuremath{\Hh_{\Sys\Frag}}}

\newcommand{\Ham}{\ensuremath{\mathbf{{H}}}}
\newcommand{\Hspace}{\ensuremath{\mathcal{H}}}
\newcommand{\HSspace}{\ensuremath{\mathcal{B}(\Hspace)}}

\newcommand{\Hint}{\ensuremath{\Ham_{\mathrm{int}}}}
\newcommand{\Henv}{\ensuremath{\Ham_{\mathrm{env}}}}
\newcommand{\Hsys}{\ensuremath{\Ham_{\mathrm{sys}}}}

\newcommand{\Msys}{\ensuremath{m_{\Sys}}}
\newcommand{\Menv}[1]{\ensuremath{m_{#1}}}
\newcommand{\omegasys}{\ensuremath{\omega_{\Sys}}}
\newcommand{\omegaenv}[1]{\ensuremath{\omega_{#1}}}
\newcommand{\omegabare}{\ensuremath{\Omega_0}}
\newcommand{\omegashift}{\ensuremath{\delta\Omega}}
\newcommand{\CC}[1]{\ensuremath{C_{#1}}}
\newcommand{\xsys}{\ensuremath{x_{\Sys}}}
\newcommand{\xenv}[1]{\ensuremath{y_{#1}}}
\newcommand{\psys}{\ensuremath{p_{\Sys}}}
\newcommand{\penv}[1]{\ensuremath{q_{#1}}}

\newcommand{\Inr}{\ensuremath{\I_{\mathrm{NR}}}}
\newcommand{\Ir}{\ensuremath{\I_{\mathrm{R}}}}
\newcommand{\Iq}{\ensuremath{\I_{\mathrm{Q}}}}
\newcommand{\EMConst}{\ensuremath{\gamma_{\scriptscriptstyle{\mathrm{EM}}}}}

\newcommand{\denom}{\ensuremath{\left((\Omega+\omega)^2 + \gamma^2\right)\left((\Omega-\omega)^2 + \gamma^2\right)}}
\newcommand{\twovec}[2]{\ensuremath{\left(\begin{array}{c}#1\\#2\end{array}\right)}}
\newcommand{\twomat}[4]{\ensuremath{\left(\begin{array}{cc}#1&#2\\#3&#4\end{array}\right)}}
\newcommand{\Si}{\ensuremath{\mathrm{Si}}}
\newcommand{\Ci}{\ensuremath{\mathrm{Ci}}}
\def\FCW{0.98\columnwidth}
\def\HPW{0.48\textwidth}
\def\TQPW{0.73\textwidth}
\def\FPW{0.98\textwidth}
\renewcommand{\omegasys}{\ensuremath{\omega_{\scriptscriptstyle\Sys}}}
\renewcommand{\Msys}{\ensuremath{m_{\scriptscriptstyle\Sys}}}
\newcommand{\Subsys}{\ensuremath{\mathcal{A}}}


\title{Quantum Darwinism in quantum Brownian motion: the vacuum as a witness}
\date{\today}
\author{Robin Blume-Kohout$^{1,2}$ and Wojciech H. Zurek$^1$}
\affiliation{$^1$ Theory Division, LANL, Los Alamos, NM 87545; $^2$ IQI, Caltech, Pasadena, CA 91125}
\begin{abstract}
We study quantum Darwinism -- the redundant recording of information about a decohering system
by its environment -- in zero-temperature quantum Brownian motion.  An initially nonlocal quantum state leaves a record whose redundancy increases rapidly with its spatial extent.  Significant delocalization (e.g., a Schr\"odinger's Cat state) causes high redundancy:  Many observers can measure the system's position without perturbing it. This explains the objective (i.e. classical) existence of einselected, decoherence-resistant pointer states of macroscopic objects.
\end{abstract}


\maketitle

A quantum system ($\Sys$) decoheres when monitored by its environment ($\Env$) \cite{PazLH01} \cite{ZurekRMP03}. That environment can act as a ``witness'', recording information about $\Sys$.
When many copies exist, the information is \emph{redundant}, and effectively objective: many observers can obtain it, but no one can change or erase it.  Objective existence is a defining feature of classical reality.  When information about one observable is redundant, information about complementary observables becomes inaccessible and it effectively ceases to exist 
\cite{ZurekRMP03,OllivierPRL04,RBKPRA06}.  This selective proliferation of ``fit'' information, at the expense of incompatible (complementary) information, is \emph{quantum Darwinism}.  

In this paper, we demonstrate quantum Darwinism in zero temperature quantum
Brownian motion (QBM). A harmonic oscillator system ($\Sys$) evolves in contact 
with a bath ($\Env$) of harmonic oscillators.  We focus on the macroscopic regime,
where the system is massive and underdamped.  In this limit, we show how redundancy
increases with the spatial extent of system's wavefunction, so that many fragments
of $\Env$ ``know'' the location of $\Sys$.

To study how information about $\Sys$ appears redundantly in $\Env$
during decoherence we must analyze the state of $\Env$, not trace it out.
In this ``environment as a witness'' paradigm, $\Env$ is not a sink
for information, but a resource from which it might be extracted.  Quantum Darwinism was introduced recently (see \cite{ZurekRMP03} and references therein), and
investigated in \cite{OllivierPRL04}.  Here, we
pursue the formulation of \cite{RBKPRA06}.

\begin{figure}[tb]
\includegraphics[width=\FCW]{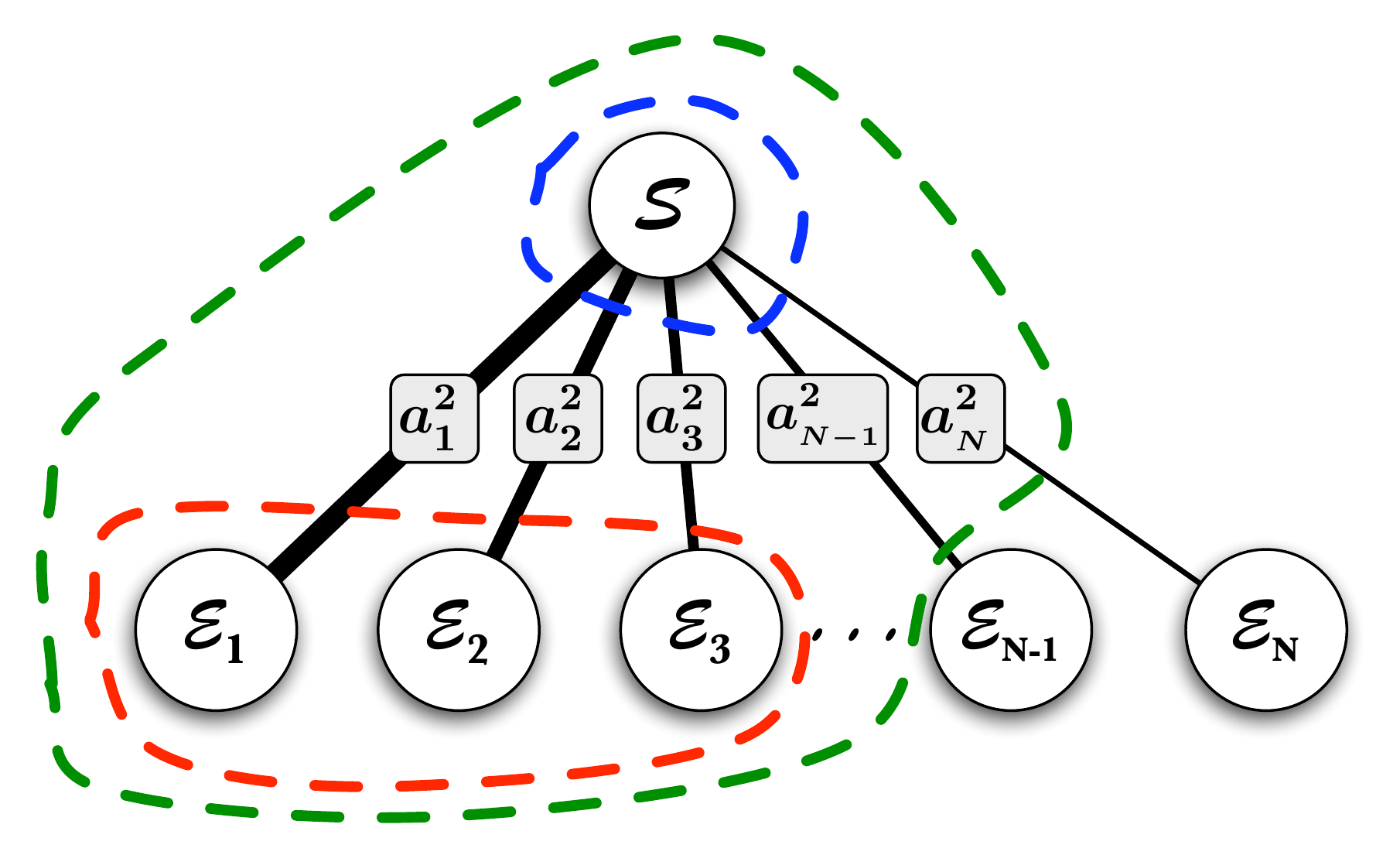}
\caption{Information about the system can be extracted from fragments -- collections of environment
subsystems. In QBM, in the weak-dissipation limit, evolved states of $\Sys$ and $\Env$ reflect the structure of the interaction Hamiltonian.  Each band $\Env_\omega$ of $\Env$ develops independent correlations with $\Sys$ (black lines), quantified by extra \emph{squared symplectic area} ($a^2_\omega$) induced in $\Sys$ and $\Env_\omega$.  
A fragment $\Frag$ (red) comprises several (not necessarily contiguous) bands.  $\Sys$ itself (blue) and the joint $\Sys\otimes\Frag$ (green) are also fragments.  Symplectic area is approximately additive, so $a^2_\Frag$ is a sum over edges connected to $\Frag$.  We use $a^2$ to compute entropy, and thence mutual information $\I_{\Sys:\Frag}$.}
\label{figStructure}
\end{figure}

The core question is ``How much information about $\Sys$ can an observer extract
from $\Env$?''  $\Env$ consists of \emph{subenvironments} $\Env_i$ ($\Env$ = $\Env_1 \otimes \Env_2 \otimes \Env_3\ldots$). Each observer has exclusive access to a fragment $\Frag$ comprising $m$ subenvironments (see Fig. \ref{figStructure}).  We factor the QBM bath into its component oscillators or \emph{bands}.  This fixed decomposition, which breaks unitary invariance and is justified by $\Env$'s interaction with apparatus, is essential \cite{RBKPRA06}.

We measure ``information'' by the quantum mutual information between $\Sys$ and $\Frag$,
\begin{equation}
\Isf = \Hs + \Hf - \Hsf,
\end{equation}
where $\Hh$ is the von Neumann entropy of a reduced density matrix.  $\Isf$
is an upper bound for the entropy (in $\Sys$) eliminated
by measuring $\Frag$.  The bound is tight for classical correlations, but quantum
correlations raise $\Isf$ above classically-allowed values. This \emph{quantum discord} 
\cite{OllivierPRL02} represents the ability to choose between
several non-commuting observables (e.g., of $\Sys$). 
In presence of decoherence (inflicted on the $\Sys \Frag$ pair by the rest of $\Env$) discord is expected to be small \cite{ZurekRMP03,OllivierPRL02}.

\begin{figure*}[tb!]
\setlength{\unitlength}{\textwidth}
\begin{picture}(1,0.4)
\put(0,0){\includegraphics[width=0.6\textwidth]{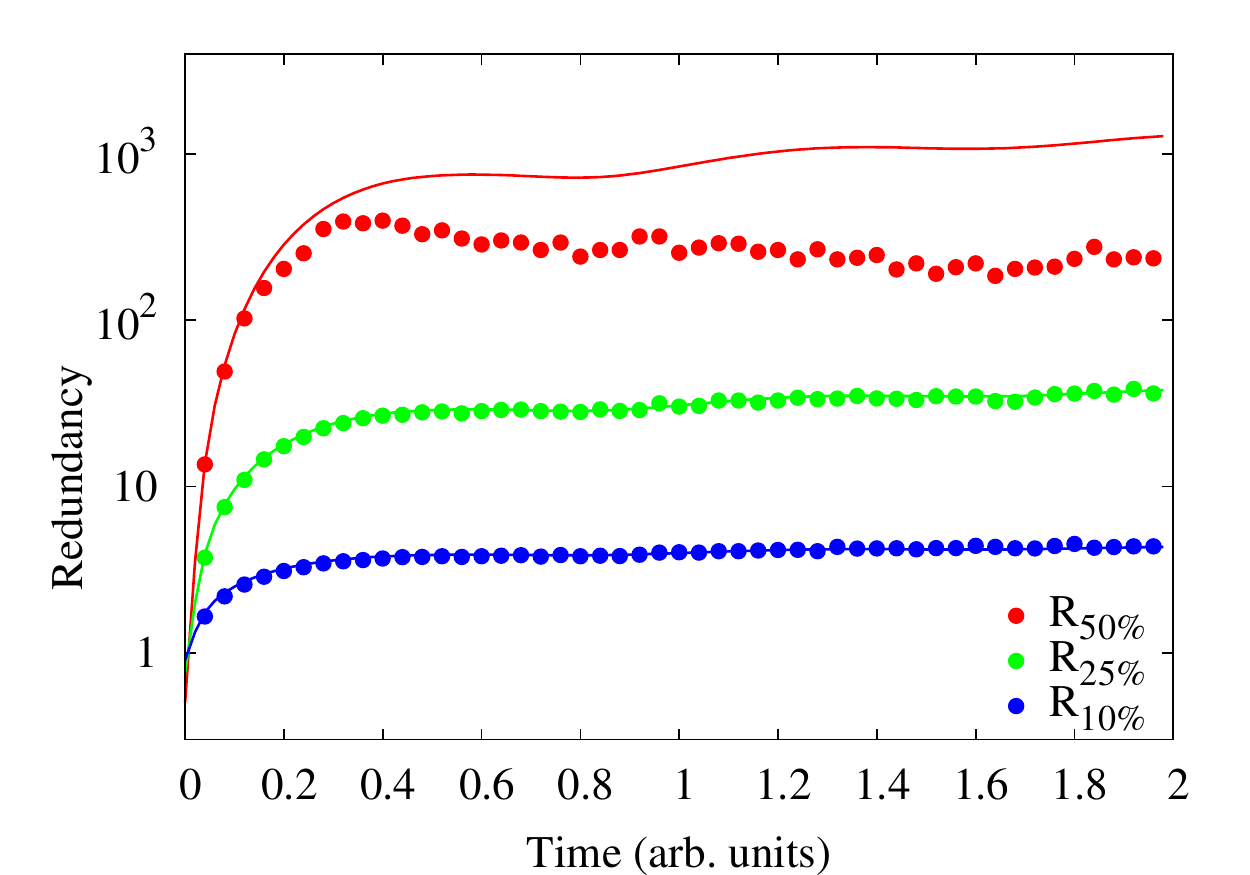}} \put(0.52,0.37){\mbox{\large\textbf{(a)}}}
\put(0.575,0.275){\includegraphics[width=0.42\textwidth]{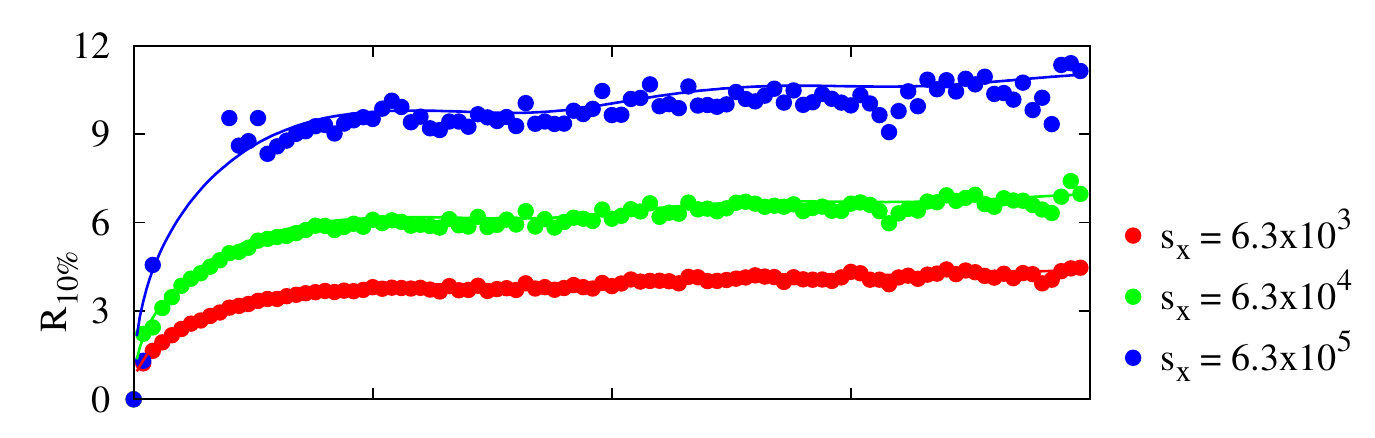}} \put(0.91,0.375){\mbox{\large\textbf{(b)}}}
\put(0.575,0.14){\includegraphics[width=0.42\textwidth]{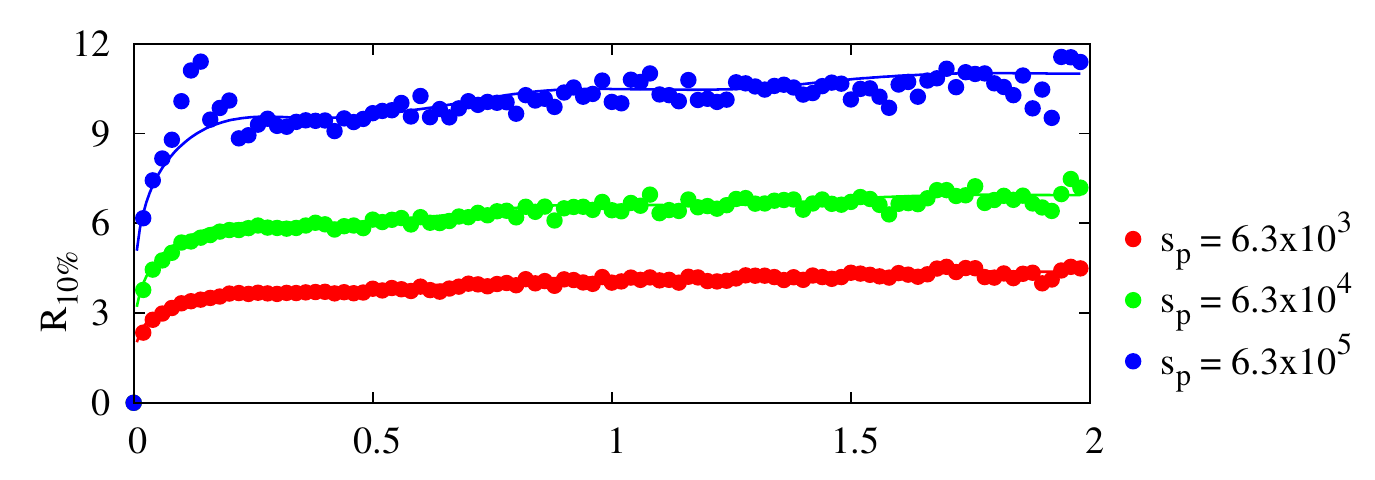}} \put(0.91,0.253){\mbox{\large\textbf{(c)}}}
\put(0.575,0){\includegraphics[width=0.42\textwidth]{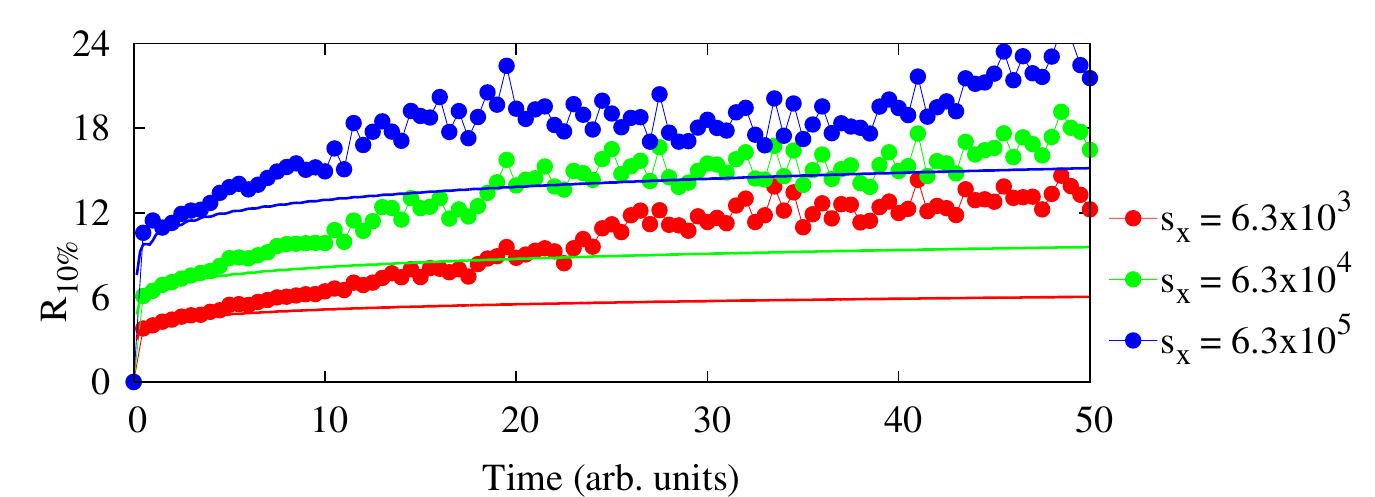}} \put(0.91,0.12){\mbox{\large\textbf{(d)}}}
\end{picture}
\caption{\emph{Delocalized states of a decohering oscillator ($\Sys$) are redundantly recorded by the environment ($\Env$).}  Plot \textbf{(a)} shows redundancy ($R_\delta$) vs. imprecision ($\delta$), when $\ket{\psi(0)}$ is squeezed in $x$ by $s_x = 6.3\times10^3$.  Plots \textbf{(b-d)} show $R_{10\%}$ -- redundancy of 90\% of the available information -- vs. initial squeezing ($s_x$ or $s_p$).  Dots denote numerics; lines -- our theory.
\emph{Details:} $\Sys$ has mass $\Msys=1000$, $\omegasys=4$.  $\Env$ comprises oscillators with $\omega\in[0\ldots16]$ and mass $\Menv{}=1$.  The frictional (coupling) frequency is $\gamma=\frac{1}{40}$.
\emph{Discussion:}  Redundancy develops with decoherence: $p$-squeezed states [plot \textbf{(c)}] decohere almost instantly, while $x$-squeezed states [plot \textbf{(b)}] decohere as a $\frac\pi2$ rotation transforms them into $p$-squeezed states.  Redundancy persists thereafter [plot \textbf{(d)}]; dissipation intrudes by $t\sim O(\gamma^{-1})$, causing $R_{10\%}$ to rise above our simple theory.  
Redundancy increases \emph{exponentially} -- as $R_\delta \approx s^{2\delta}$ -- with imprecision [plot \textbf{(a)}]. So, while $R_{\delta} \sim 10$ may seem modest, $\delta=0.1$ implies \emph{very} precise knowledge (resolution around 3 ground-state widths) of $\Sys$.
This is half an order of magnitude better than a recent record \cite{LahayeScience04} for measuring a micromechanical oscillator.  At $\delta \sim 0.5$ -- resolving $\sim \sqrt{s}$ different locations within the wavepacket -- $R_{50\%}\gtrsim 10^3$ (our maximum numerical resolution).
}
\label{figRedundancy}
\end{figure*}
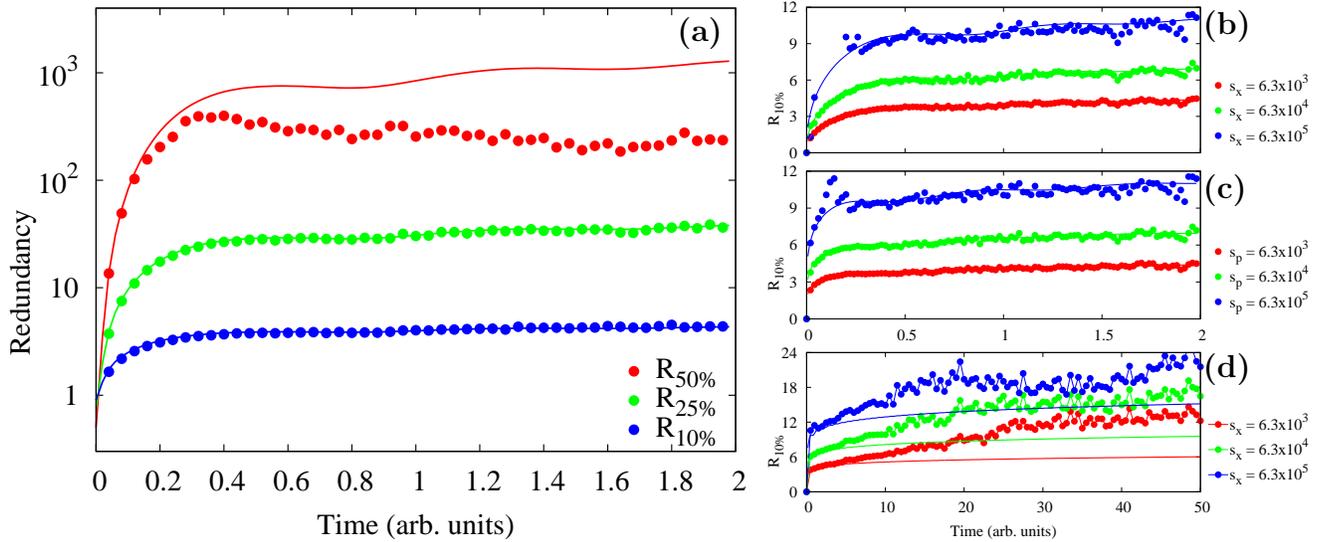

\begin{figure}[tb]
\begin{tabular}{l}
\vspace{-0.15in} \includegraphics[width=\FCW]{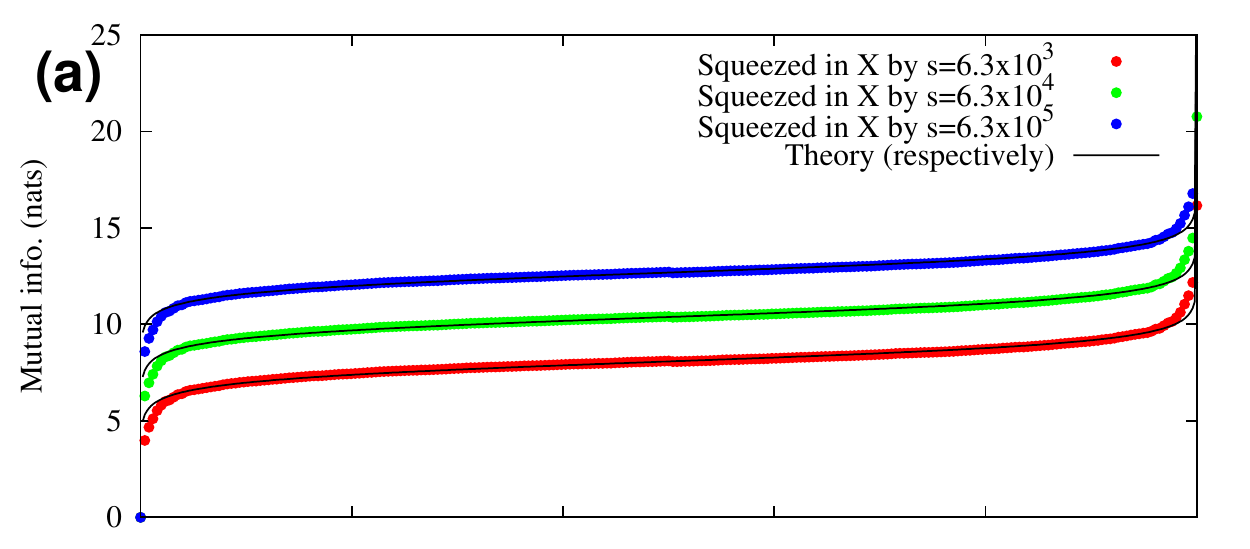}\\
\includegraphics[width=\FCW]{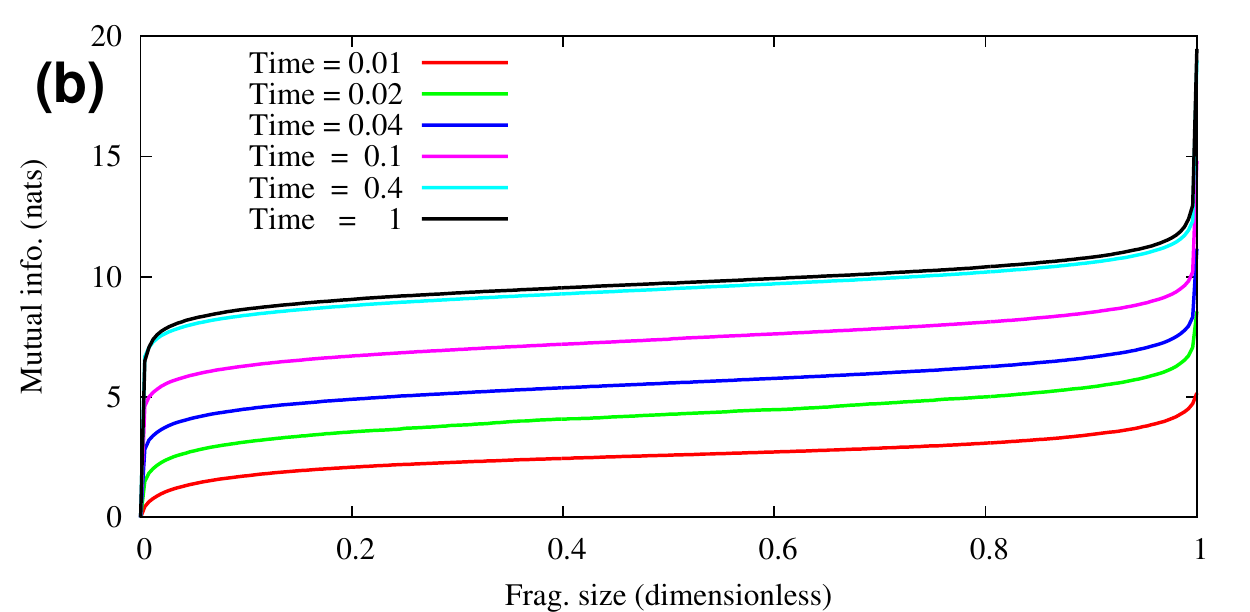}\\
\end{tabular}
\caption{\emph{Partial information plots} (PIPs) show how information is stored in $\Env$.  They illustrate how $\Isf$ -- the amount of information in a randomly chosen fragment $\Frag\subset\Env$ -- depends on $\Frag$'s size.  Here, we initialized $\Sys$ in an $x$-squeezed state, which decoheres as it evolves into a superposition of $x$ states.  Plot \textbf{(a)} shows PIPs for three fully-decohered ($t=4$) states with different squeezing.  Small fragments of $\Env$ provides most of the available information about $\Sys$; squeezing changes the amount of \emph{redundant} information without changing the PIP's shape.  The numerics agree with a simple theory.  Plot \textbf{(b)} tracks one state as decoherence progresses.  Again, PIPs' shape is invariant; time only changes the amount of redundant information.}
\label{figPIPs}
\end{figure}

We use two tools to analyze information storage. \emph{Partial information} is the average information in a random fragment
containing a fraction $f$ of $\Env$,
\begin{equation}
\Ibar(f) = \mathrm{avg}_{\mathrm{all\ } \Frag \mathrm{\ of\ size\ } f}\left(\Isf\right),
\end{equation}
\emph{Partial information plots} (PIPs) assume a characteristic shape in the presence of
redundancy: $\Ibar(f)$ increases sharply around $f=0$ and $f=1$, but has a long,
flat ``classical plateau'' in between. Thus, almost all (all but $\delta$) of
this classical information can be extracted from a small fraction $f_\delta$ of $\Env$. 
\emph{Redundancy} ($R_\delta$) is just the number 
of disjoint fragments $\Frag$ that provide all but $\delta$ of the available
information about $\Sys$ -- i.e., satisfying $\Isf \geq (1-\delta)\Hs$, or;
\begin{equation}
R_\delta=\frac 1 {f_\delta}
\end{equation}
Further discussion of $R_\delta$ and PIPs 
(see Figs. \ref{figRedundancy}, \ref{figPIPs}), 
is found in \cite{RBKPRA06}.

The QBM \cite{FeynmanAOP63,CaldeiraPhysicaA83,UnruhPRD89,HuPRD92} Hamiltonian
\begin{equation}
\Ham = \Hsys + \frac12
      \sum_{\omega}{\left(\frac{\penv{\omega}^2}{\Menv{\omega}}
     + \Menv{\omega}\omega^2 \xenv{\omega}^2\right)}
+ \xsys\sum_{\omega}{\CC{\omega}\xenv{\omega}} \label{eqQBMHamiltonian}.
\end{equation}
describes a central oscillator whose position $\xsys$ is linearly coupled to a bath of oscillators.  The central system obeys $\Hsys = (\frac{\psys^2}{\Msys} + \Msys\omegabare^2 \xsys^2)/2$; the environmental coordinates $\xenv{\omega}$ and $\penv{\omega}$ describe a single band (oscillator) $\Env_\omega$.  
As usual, the bath is defined by its spectral density,
$I(\omega) = \sum_{n}{\delta\left(\omega-\omegaenv{n}\right) 
            \frac{\CC{n}^2}{2\Menv{n}\omegaenv{n}}}, \label{eqSpectralDensity}$
which quantifies the coupling between $\Sys$ and each band of $\Env$.  We consider an
\emph{ohmic} bath with a cutoff $\Lambda$ (see note
\footnote{We adopt a sharp cutoff (rather than the usual smooth rolloff) to simplify numerics.}):
$I(\omega) = \frac{2\Msys\gamma_0}{\pi}\omega$
for $\omega\in[0\ldots\Lambda]$.
Each coupling is a differential element,
$\diff\CC{\omega}^2 = \frac{4\Msys\Menv{\omega}\gamma_0}{\pi}\omega^2\diff\omega$
for $\omega\in[0\ldots\Lambda]$.
For numerics, we divide $[0\ldots\Lambda]$ into discrete
bands of width $\Delta\omega$, which approximates the exact model
well up to a time $\tau_{rec} \sim
\frac{2\pi}{\Delta\omega}$.


We initialize $\Sys$ in a squeezed coherent state, and $\Env$ in its ground state.
QBM's linear dynamics preserve the Gaussianity of the initial state, which
can be described by its mean and variance:
\begin{equation}
\vec{z} = \left(\begin{array}{c}\expect{x}\\ \expect{p}\end{array}\right)\mbox{\ ;\ }
\Delta = \left(\begin{array}{cc}\Delta x^2 & \Delta xp \\
 \Delta xp & \Delta p^2\end{array}\right). 
\label{eqVariance}
\end{equation}
Its entropy, $\Hh(\rho) = -\Tr\rho\ln\rho$, is a function of its \emph{squared symplectic area},
\begin{eqnarray}
&a^2 = \left(\frac{\hbar}{2}\right)^{-2}\det(\Delta) &\label{eqSquaredAreaDef} \\
&\Hh(a) = \frac12\left(\begin{array}{l}\ (a+1)\ln(a+1)\\-(a-1)\ln(a-1)\end{array}\right)-\ln2
\approx \ln\left(\frac{e}{2}a\right),& \label{eqGaussianEntropy}
\end{eqnarray}
where $e$ is Euler's constant, and the approximation 
is excellent for $a>2$.  For multi-mode states, numerics yield $\Hh(\rho)$ exactly as a sum
over $\Delta$'s symplectic eigenvalues \cite{SerafiniJPB04}, but our theoretical treatment approximates a collection of oscillators as a single mode with a single $a^2$.

Exact solutions to QBM, even for the reduced dynamics of $\Sys$ alone, are nontrivial.  Quantum Darwinism requires a more extensive solution describing the dynamics of $\Env$.
We obtain it numerically, describing the initial Gaussian product state with a covariance matrix (Eq. \ref{eqVariance}), evolving it by canonical methods (see \cite{AnglinMPLA96,RBKPRA03}), and computing mutual information from symplectic area.
To compute redundancy ($R_\delta$), we apply a Monte Carlo technique to find the
amount of randomly selected bandwidth required to obtain $\Isf = (1-\delta)\Hs$.
We choose units where: $\frac{\hbar}{2}=1$; the masses of the $\Env_\omega$ are 1;
the renormalized frequency of $\Sys$ is 4; and the bath frequencies lie in $[0,\Lambda=16]$.
The frictional coefficient $\gamma_0$ varies with $\Msys$ so that $\Msys\gamma_0=25$; most often, $\Msys=10^3$ and $\gamma_0=\frac{1}{40}$.

Our main result is that substantial redundancy appears in the QBM model (Fig. 
\ref{figRedundancy}).  Redundancy depends on the initial squeezing $s$, so that $R_{\delta} \sim s^{2 \delta}$.
It appears along with decoherence -- rapidly for $\hat{p}$-squeezed
states (Fig. \ref{figRedundancy}b), more slowly for $\hat{x}$-squeezed states
(Fig. \ref{figRedundancy}a) \footnote{$x$-squeezed states are extended in $p$, and decohere as the system's dynamics rotate $p$ into $x$, which then decoheres.
} -- then remains relatively constant.  However, dissipation (not
analyzed here) causes redundancy to further increase on a timescale
$t\sim O(\gamma_0^{-1})$ (see Fig. \ref{figRedundancy}d).

PIPs (Fig. \ref{figPIPs}) show how information about $\Sys$ is stored in $\Env$.
$\Ibar(f)$ rises rapidly as the fragment's size ($f$) increases from zero,
then flattens for larger fragments.  
Most -- all but $\sim1$ nat -- of $\Hs$ is redundant.
When $\Sys$ is macroscopic, this \emph{non-redundant information}
is dwarfed by the total amount of information lost to $\Env$.

Let us now derive a model for
this behavior.  Suppose $\Sys$ is macroscopic, so $\Msys\rightarrow\infty$.  The bath's
spectral density is independent of $\Msys$, so $\Msys\gamma_0$ remains constant, and
$\gamma_0$ is small.  The mutual information between $\Sys$ and a fragment $\Frag$
depends on the entropies of $\rho_\Sys$, $\rho_\Frag$, and $\rho_{\Sys\Frag}$, so we compute their squared symplectic areas.

As $\Msys\rightarrow\infty$, the kinetic term in $\Hsys$
(Eq. \ref{eqQBMHamiltonian}) becomes insignificant. $\Hsys$ thus commutes with the
interaction term, and can be ignored.  The remainder of $\Ham$ has the form
$\Id_\Sys \otimes \sum_\omega{\Ham_\omega} + \hat\xsys \otimes \sum_\omega{\R_\omega}$.
When $\ket{\psi_\Sys} = \ket{x}$, each $\Env_\omega$ feels a
well-defined $\Ham_\omega(x)$, and evolves as 
$\ket{\psi_\omega(0)}\rightarrow\ket{\psi_\omega(t;x)}$, \emph{conditional} upon the value
of $x$.  When $\ket{\psi_\Sys(0)}$ is a superposition of $\ket{x}$ states, the 
product state evolves into a Gaussian \emph{singly-branching state} \cite{RBKPRA06};
\begin{eqnarray}
& \left(\int{\!\psi_\Sys(x)\ket{x}\diff x}\right) \otimes \ket{\psi_1(0)}\ket{\psi_2(0)}\ldots\ket{\psi_{\Nenv}(0)}& \\
& \Downarrow \nonumber & \\ 
& \int{\psi_{\Sys}(x) \ket{x}\otimes\ket{\psi_1(t;x)} \ket{\psi_2(t;x)}\ldots\ket{\psi_{\Nenv}(t;x)}\diff{x}},
\label{eqQBMBranchingStates}
\end{eqnarray}
The reduced state $\rho_\Subsys$ for any
subsystem $\Subsys$ is spectrally equivalent to a partially-decohered state of $\Sys$:
\begin{equation}
\rho_\Subsys(x,x') = \rho_\Sys(x,x',t=0)\Gamma_\Subsys(x,x').
\end{equation}
The decoherence factor $\Gamma_\Subsys(x,x')$ is a product (over all $\Env_\omega$ \emph{not}
in $\Subsys$ if $\Subsys$ contains $\Sys$; otherwise, over all $\Env_\omega$ in $\Subsys$) of contributions 
$\Gamma_{\omega}(x,x') \equiv \braket{\psi_\omega(t;x)}{\psi_\omega(t;x')}$ from individual
bands.

$\Gamma_{\omega}(x,x')$ measures a band's power to decohere $\ket{x}$ from $\ket{x'}$.
Let us define an \emph{additive} decoherence factor $d\propto\log\Gamma$.  The logarithm is always proportional to $(x-x')^2$ (see Eq. \ref{eqQBMGammaSq}), so we set
\begin{equation}
d_{\omega}(t) \equiv -\frac{\log\left(\Gamma_{\omega}(x,x')\right)}{(x-x')^2}. \label{eqQBM_dfactor}
\end{equation}
For a continuous spectral density, $d_{\omega}$ is a differential
$\diff d_{\omega} = \frac{\diff d}{\diff\omega}\diff\omega$, and the decoherence $d_{\Subsys}$
experienced by a subsystem $\Subsys$ is an integral over its bandwidth.

Suppressing off-diagonal elements of $\rho$ affects $\hat{x}$ not at all, but 
increases $\Delta p^2$ by $\delta p^2_\Subsys = 2\hbar d_{\Subsys}$, so
\begin{equation}
a^2_\Subsys \rightarrow 1 + \delta a^2_\Subsys = 1 + \left(\frac{\hbar}{2}\right)^{-2}\Delta x^2\delta p^2_\Subsys = 1 + \frac{8\Delta x^2}{\hbar}d_\Subsys.
\end{equation}
This $\delta a^2$ is a key quantity.  It measures the correlation-induced uncertainty in $\Subsys$ and its
complement, and therefore 
the amount of correlation.
For example, the correlation between $\Sys$ and $\Env$ is the uncertainty in $\Sys$, given by an integral over all bands of $\Env$:
$\delta a^2_\Sys = \frac{8\Delta x^2}{\hbar}\int_{0}^{\Lambda}{\frac{\diff d}{\diff\omega}\diff\omega}.$
The uncertainty in a fragment $\Frag$ is the integrated $\diff a^2$ from all its component bands; that in $\Sys\Frag$ is the integrated $\diff a^2$ for its complement,
$\overline{\Frag}$ (where $\Env=\Frag\otimes\overline{\Frag}$; see Fig. \ref{figStructure}).

\begin{figure}[tb]
\begin{tabular}{l}
\vspace{-0.15in} \includegraphics[width=\FCW]{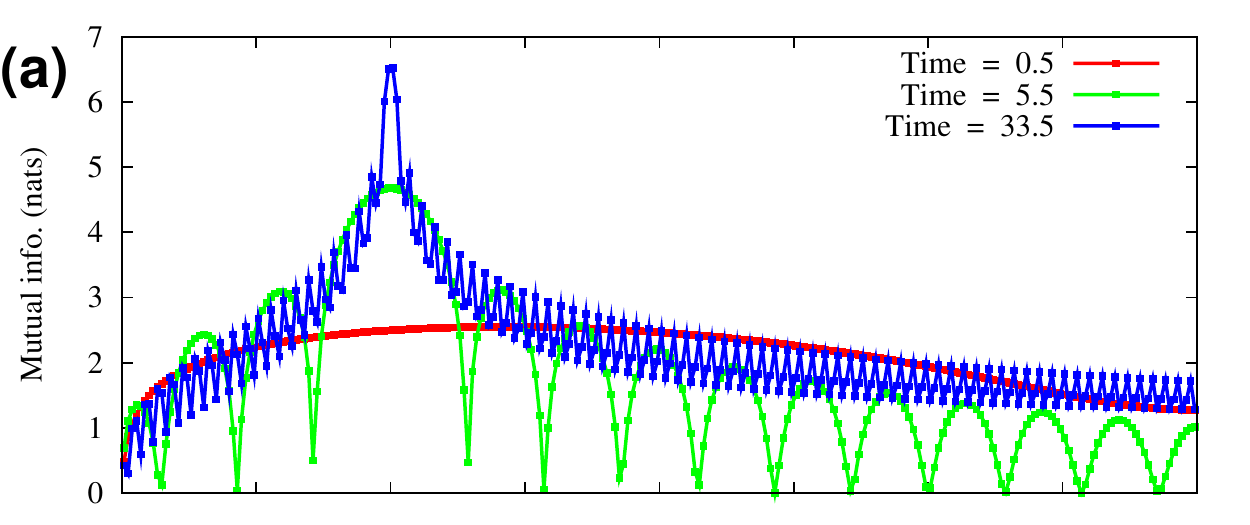}\\
\includegraphics[width=\FCW]{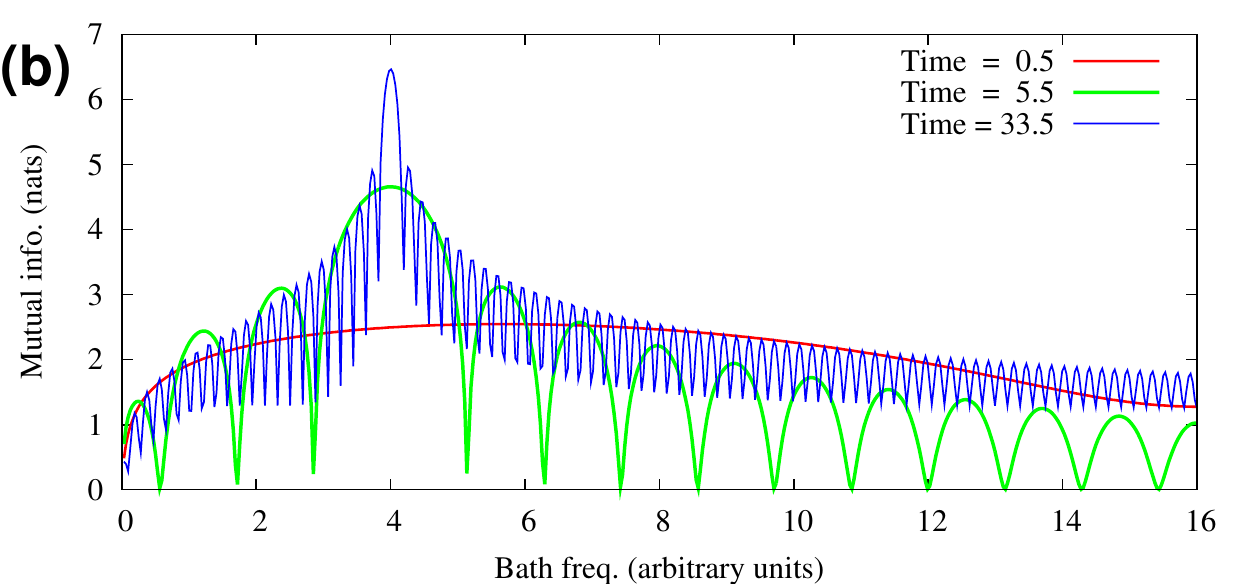}\\
\end{tabular}
\caption{\emph{Different bands of $\Env$ hold different amounts of information.}  Here, $\Sys$ is prepared with a squeezing of $s=6.3\times10^3$, and the dissipation constant is $\gamma = \frac{1}{400}$.  Plot \textbf{(a)} shows numerics, while \textbf{(b)} shows theory (see Eq. \ref{eqDrivenD}).  Initially (red), all bands participate.  Later (green), resonant bands around $\omega\sim\Omega=4$ become more important.  After many oscillations (blue), resonant bands dominate.  Theory agrees extremely well, though small discrepancies appear later.}
\label{figMutual}
\end{figure}

When $\Sys$ is in state $\ket{x}$, $\Env_\omega$ experiences a Hamiltonian
\begin{equation}
\Ham_\omega(x) = \frac{\penv{\omega}^2}{2\Menv{\omega}} + \frac{\Menv{\omega}\omega^2}{2}
    \left(\xenv{\omega} - \delta\xenv{\omega}\right)^2
	- \frac{\Menv{\omega}^2 \omega^2}{2} \delta\xenv{\omega}^2. \label{eqConditionalH}
\end{equation}
Its initial (ground) state $\ket{\psi_\omega(0)}$ evolves into a coherent state $\ket{\psi_\omega(t;x)}$, along a circle of radius $\delta\xenv{\omega} = C_\omega/(\Menv{\omega}\omega^2)$.
Solving the equation of motion and inserting $\Delta\xenv{0}^2 = \hbar/2 \Menv\omega$
and $\Delta\penv{0}^2 = \hbar\Menv\omega/2$ yields
\begin{equation}
\left|\Gamma\idx{\omega}_{x,x'}\right| = 
     \exp\left[-\frac{\CC{\omega}^2}{2\Menv{\omega}\hbar\omega^3}(x-x')^2
     \left(1-\cos\omega t\right)\right]. \label{eqQBMGammaSq}
\end{equation}
The exponent is (as promised) proportional to $(x-x')^2$, and 
$\diff d_\omega = \frac{2\Msys\gamma_0}{\pi\hbar\omega}(1-\cos\omega t)\diff\omega.$

Beyond $t\sim O(\omegasys^{-1})$, $\Hsys$ becomes
relevant.  $\Env_\omega$ is driven, not just displaced, by $\Sys$.  
$\Sys$ is very massive, so it acts as a classical driving force on $\Env_\omega$.
To model this, we substitute $x=x_0\cos\omegasys t$ into Eq. 
\ref{eqConditionalH} and re-solve the ensuing equation of motion to get
\begin{equation}
\frac{\diff d}{\diff\omega} = \frac{\Msys\gamma_0}{\pi\hbar}\frac{\omega^3\diff\omega}{(\omegasys^2-\omega^2)^2}
\left[\begin{array}{l}\ \ \left(\sin\omega t - \frac{\omegasys}{\omega}\sin\omegasys t\right)^2 \vspace{0.05in} \\ + (\cos\omega t - \cos\omegasys t)^2\end{array}\right]. \label{eqDrivenD}
\end{equation}
Integrating over $\omega$ yields a cumbersome formula for $\delta a^2_\Sys$, and thus for $\Hs(t)$.

We can now predict PIPs ($\Ibar(f)$).  When $\Frag$ contains a 
randomly selected fraction $f$ of $\Env$'s bandwidth, $\rho_{\Frag}$'s squared area is 
$a^2_{\Frag} = 1 + f\delta a^2_\Sys$, and that of $\rho_{\Sys\Frag}$ is
$a^2_{\Sys\Frag} = 1 + (1-f)\delta a^2_\Sys$.  Applying Eq. (\ref{eqGaussianEntropy}) (where $\delta a^2_\Sys \gg 1$) yields
\begin{equation}
\Isf \approx \Hh_\Sys + \frac12\ln\left(\frac{f}{1-f}\right) \label{eqApproxPIP}.
\end{equation}
This simple result fits numerics very well (pre-dissipation), and predicts the shape-invariance of PIPs.

We can also predict \emph{where} information is stored in $\Env$.  
If $\Env_\omega$ is a band at frequency $\omega$, of width $\Delta\omega$, then $\I_{\Sys:\Env_\omega} = \Hh(\Sys)+\Hh(\Env_\omega) - \Hh(\Sys\Env_\omega) \approx \Hh(\Env_\omega)$.  The band's entropy is computed from its decoherence factor, $d_{\Env_\omega} \approx \Delta\omega\frac{\diff d}{\diff \omega}$ (Eq. \ref{eqDrivenD}).  The results agree with numerics (Fig. \ref{figMutual}).

Redundancy counts the number of disjoint fragments with
$\Isf \geq (1-\delta)\Hs$.  Because $\Isf$ depends only on the fragment's size ($f$), 
 $\Isf \geq (1-\delta)\Hs$ iff 
$f \geq f_\delta = \frac{e^{-2\delta\Hs}}{1+e^{-2\delta\Hs}}$.
$\Env$ contains $N_\delta = 1/f_\delta$ such fragments, so
\begin{equation}
R_\delta \approx e^{2\delta\Hs} \approx s^{2\delta}. \label{eq17}
\end{equation}
The second equality follows because an $s$-squeezed state decoheres to a mixed state with $H_\Sys \approx \ln s$.  Eq. (\ref{eq17}) is a succinct and easy-to-use summary of our results, and fits the data well (see Fig. \ref{figRedundancy}). For instance, at $\delta=0.5$, we localize $\Sys$ with accuracy $\sim\sqrt s$, with redundancy $R_{0.5} \propto s$ (see Fig. \ref{figRedundancy}c). 

To generalize this result, observe that squeezing controls the initial \emph{spatial extent} ($\Delta\xsys$), and that redundancy increases rapidly with $\Delta\xsys$.  A fragment of $\Env$ provides a fuzzy measurement of $\Sys$ (whose resolution increases with its size).  A Schr\"odinger's Cat state will yield high redundancy (but only $\sim1$ bit of entropy), because small fragments are sufficient to resolve the two branches.


We have provided convincing evidence for quantum Darwinism in one of the most-studied models of decoherence.  Our theory of the $\Sys-\Env$ information flows, using singly-branching states, effectively models detailed numerics, and leads to a compelling picture:  redundancy (e.g., Eq. \ref{eq17}) accounts for objectivity and classicality; the environment is a witness, holding many copies of the evidence.  Though we did not discuss dissipation (which requires more sophisticated analysis), it actually \emph{increases} $R_\delta$, by reducing \emph{non}-redundant correlations.  We postpone discussion of quantum Darwinism in the dissipative regime, and comparisons with the case of discrete pointer observables, to forthcoming papers.

We acknowledge stimulating discussions and useful comments on the manuscript by David Poulin.



\bibliographystyle{apsrev}
\bibliography{/home/rbk/bib/decoherence,/home/rbk/bib/quantum,/home/rbk/bib/QBM,/home/rbk/bib/RBK,/home/rbk/bib/Zurek,/home/rbk/bib/math}
\end{document}